\documentclass[11pt,reqno]{amsart}
\usepackage{graphicx}
\usepackage{mathtools}
\usepackage{ulem}
\usepackage{color}

\usepackage{amsmath}
\usepackage{amssymb}
\usepackage{graphicx}
\usepackage{accents}
\usepackage{pstricks,pst-node,pst-coil,pst-plot}
\usepackage{mathtools}
\usepackage{ulem}
\usepackage{mathtools}
\usepackage{ulem}
\usepackage{graphicx}
\usepackage{color}
\usepackage{graphicx}
\usepackage{color}
\usepackage{dcolumn}
\usepackage{graphicx}
\usepackage{color}
\usepackage{dcolumn}
\usepackage{bm}
\usepackage{mathrsfs}
\usepackage{graphicx}
\usepackage{color}
\usepackage{dcolumn}
\usepackage{bm}
\usepackage{slashed}
\usepackage{epstopdf}
\usepackage{slashed}
\usepackage{bm}
\usepackage{slashed}
\usepackage{dcolumn}
\usepackage{bm}
\usepackage{slashed}
\usepackage{color}
\usepackage{color}

\newtheorem{case}{Example}
\newtheorem*{thm}{Theorem}
\newcommand{\be}{\begin{equation}}
\newcommand{\ee}{\end{equation}}
\def\mh{E_{\mbox{Hawking}}}
\def\eg{E_{\mbox{Geroch}}}

\def\mhy{E_{\mbox{Hayward}}}
\def\Area{|\Sigma_t|}

\newtheorem*{coro}{Corollary}

\def\ringA{\accentset{\circ}{A}}
\newcommand{\beq}{\begin{equation}}
\newcommand{\eeq}{\end{equation}}
\newcommand{\bee}{\begin{equation*}}
\newcommand{\eee}{\end{equation*}}
\def\d{\mathrm{d}}
\def\mhy{E_{\mbox{Hayward}}}

\let\<\langle
\let\>\rangle

\newcommand{\definedas}{\mathrel{\raise.095ex\hbox{\rm :}\mkern-5.2mu=}}

\begin{document}


\title[Quasi-local energy and Oppenheimer-Snyder collapse]{Quasi-local energy and Oppenheimer-Snyder collapse}
\author{Xiaokai He}
\address[Xiaokai He]{School of Mathematics and Computational Science, Hunan First Normal University, Changsha 410205, China}
\email{sjyhexiaokai@hnfnu.edu.cn}
\author{Naqing Xie}
\address[Naqing Xie]{School of Mathematical Sciences, Fudan
University, Shanghai 200433, China.}
\email{nqxie@fudan.edu.cn}
\dedicatory{Dedicated to the memory of Professor Chaohao Gu.}
\begin{abstract}
We consider a scenario of the gravitational collapse of the Oppenheimer-Snyder dust cloud with spatially constant matter density. Given a closed two-surface within the star, three versions of the quasi-local energy are investigated. We show that, during the collapse, (i) the Geroch energy of the surface is nonpositive and increases to zero; (ii) the Hawking energy of the surface is monotonically increasing and approaches to the Hayward energy of the initial surface; and (iii) the Hayward energy of the surface is conserved and remains positive. These results have no restrictions on the topology and the symmetry of the surface.
\end{abstract}

%




\maketitle



\section{Introduction}\label{S1}
It is widely believed that a star will collapse when it has exhausted all thermonuclear sources of energy. The occurrence of the singularity can be viewed as the consequence of the competition between the blow-up of the spacetime curvature and the shrinking of the star. In 1939,  Oppenheimer and Snyder studied the collapse of a spherically symmetric dust cloud by using a special class of Tolman solutions \cite{OS39}. The star is assumed to have uniform matter density. There is no shell-crossing and the dust shells all arrive at the same time at the centre. The time of the occurrence of the singularity is determined by the initial matter density. The greater the density is, the sooner the singularity appears.

We are given a closed two-surface in a time slice within the star. The surface will collapse and ultimately meet the singularity. As a quasi-local description of the gravitational collapse, it is natural to ask how the energy of the domain enclosed by the surface behaves. Does the energy sufficiently balance the shrinking of the surface?

There are many notions of energy in the literature (e.g. the Arnowitt-Deser-Misner energy \cite{ADM61}, the Brown-York energy \cite{BY93}, the Geroch energy \cite{Ge73}, the Hawking energy \cite{Ha68}, the Hayward energy \cite{Ha94}, the Kijowski energy \cite{Kij97}, and the Wang-Yau energy \cite{WY09}). For a review and a more detailed discussion of the various energy expressions, see e.g. \cite{Sz09}.

 In this paper, three versions of the quasi-local energy are investigated. We show that, during the collapse, (i) the Geroch energy of the surface is nonpositive and increases to zero; (ii) the Hawking energy of the surface is monotonically increasing and approaches to the Hayward energy of the initial surface; and (iii) the Hayward energy of the surface is conserved and remains positive. Even though the spacetime we consider is spherically symmetric, there are no restrictions on the topology and the symmetry of the surface in question.

The order of the paper is as follows. In Section \ref{S2}, we give a brief review of the Oppenheimer-Snyder dust model. Section \ref{S3} includes the main results and the proof. Conclusions and discussions are in the last section.

Throughout this paper, we make use of the gravitational system of units with $c=G=1$. The signature of the spacetime metric is assumed to be $(-,+,+,+)$. Unless otherwise stated, Greek indices are used to label spacetime dimensions, lower case Latin indices are reserved for three-dimensional objects, and capital ones are for two-dimensional objects.

\section{Review of the Oppenheimer-Snyder collapse}\label{S2}
This section gives a quick review of the collapse of the Oppenheimer-Snyder cloud with uniform matter density. We follow the notations of Choquet-Bruhat and most of the formulae in this section can be found in \cite[Chapter IV, Section 12]{Ch09}.

The spherically symmetric spacetime metric is written as
\be\label{spme}
\tilde{g}=-\d t^2+e^{2\omega(t,r)}\d r^2+R^2(t,r)\big(\d \Theta^2+\sin^2\Theta\d \varphi^2\big)\ee
where $r,\Theta,\varphi$ are the matter comoving coordinates. Here $R(t,r)$ is the areal radius of the spherical dust shell of parameter $r$ at time $t$.

The Einstein field equation
\bee
\mbox{Ric}^{\tilde{g}}\big(\frac{\partial}{\partial t}, \frac{\partial}{\partial r}\big)=0\eee
and the conservation law
\bee
\nabla_{\mu}T^{\mu 0}=0\eee
lead to the consequence that the density $\mu(t,r)$ at time $t$ has the form
\bee
\mu(t,r)=\frac{r^2\mu_0(r)}{R^2R^\prime}.\eee
Here $\mu_0(r)$ is an arbitrary function of $r$, $\ ^\prime$ denotes the derivative with respect to the parameter $r$, and we have chosen $R(0,r)=r$. It has been shown that the Einstein equation has a first integral
\bee
\frac{1}{2}\dot{R}^2-\frac{M(r)}{R}=\frac{1}{2}\big(f^2(r)-1\big)\eee
with
\bee
M(r)=\int_0^r f(\rho)\mu(t,\rho)R^2(t,\rho)e^\omega\d\rho.\eee
Here $f(r)$ is an arbitrary function of $r$. In this paper, we solely study the simple case of marginally bound collapse with $f(r)\equiv1$. Indeed, the mass function $M(r)$ is independent of $t$ \cite[Page 99, Eqn (12.12)]{Ch09},
\be\label{mass}
M(r)=\int_0^r\mu_0(\rho)\rho^2 \d \rho.\ee
We identify $\mu_0(r)$ with the initial density and from now on assume it to be a positive constant $\mu_0(r)=\mu_0>0$.
Consequently, one has
\be\label{R-eq}
R(t,r)=r\Big(1-\sqrt{\frac{3\mu_0}{2}}\ t\Big)^{\frac{2}{3}}
\ee
and
\be\label{w-eq}
e^{\omega(t,r)}=R^\prime(t,r)=\Big(1-\sqrt{\frac{3\mu_0}{2}}\ t\Big)^{\frac{2}{3}}.
\ee
Note that under the assumption of constant matter density, $\omega(t,r)$ is independent of the parameter $r$. We may rewrite $\omega(t,r)$ as $\omega(t)$ subsequently.

Equations \eqref{R-eq} and \eqref{w-eq} indicate that the dust shells all arrive at the centre $R=0$ at the same time and the singularity occurs at time $t^\ast=\sqrt{\frac{2}{3\mu_0}}$.

The function $\omega(t)$ satisfies a nice differential equation
\be\label{magic}
(\dot{\omega})^2e^{3\omega}=\frac{2\mu_0}{3}\ee
- the blow-up of $(\dot{\omega})^2$ balancing the shrinking of $e^{3\omega}$ as $t\rightarrow t^{\ast -}$. Here $\dot{}$ denotes the derivative with respect to the time parameter $t$.

The spacetime metric \eqref{spme} can be rewritten in the following (1+3)-form as
\be
\begin{split}\label{1+3}
\tilde {g}&=-\d t^2+g_t\\
&=-\d t^2+\Big(1-\sqrt{\frac{3\mu_0}{2}}\ t\Big)^{\frac{4}{3}}
\Big(\d r^2+r^2\d \Theta^2+r^2 \sin^2\Theta \d \varphi^2 \Big) \ \ \ (r<R_b)\\
&=-\d t^2+e^{2\omega(t)}g_0
\end{split}\ee
where $g_0=\d r^2+r^2\d \Theta^2+r^2 \sin^2\Theta \d \varphi^2$ is the Euclidean metric and $R_b$ is the $r$ parameter of the outermost dust shell at $t=0$.

The induced three-metric $g_t$ on the time slice $M_t=\{t=\mbox{const.}\}$ is also flat. The extrinsic curvature $K_t$ of $M_t$ in spacetime has the form
\be\label{Kt}
(K_t)_{ij}=\frac{\dot{(e^{2\omega})}(g_0)_{ij}}{2}=(\dot{\omega})(g_t)_{ij}.\ee
indicating that $M_t$ is totally umbilical, i.e. the extrinsic curvature is proportional to the three-metric.

\section{Main results}\label{S3}
Suppose that a closed two-surface $\Sigma_t$ lies in a time slice $M_t$ within the star. It is defined by an equation of the matter comoving coordinates $F(r,\Theta,\varphi)=0$. We take the orthonormal frame $\{e_0,e_1,e_2,e_3\}$. Here $\{e_2,e_3\}$ are tangential to the surface $\Sigma_t$, $e_1$ is the outward unit normal of $\Sigma_t$ in the time slice $M_t$, and $e_0=\frac{\partial}{\partial t}$ is the future directed timelike unit normal of $M_t$ in spacetime. Denote by $H_t$ and $A_t$ the mean curvature and the second fundamental form of $\Sigma_t$ in $M_t$ respectively.

The Geroch energy quasi-local energy \cite{Ge73} of $\Sigma_t$ is defined as
\bee\label{eg}
\eg(\Sigma_t)=\frac{1}{8\pi}\sqrt{\frac{\Area}{16\pi}}\int_{\Sigma_t} (\mbox{Scal}_{\sigma_t} -\frac{1}{2}H_t^2)\d\sigma_t.\eee
Here $\sigma_t$ is the induced two-metric on $\Sigma_t$, $\Area =\int_{\Sigma_t} \d \sigma_t$ is the area of $\Sigma_t$, and $\mbox{Scal}_{\sigma_t}$ is the scalar curvature of the metric $\sigma_t$.

The Geroch energy has a very nice monotonicity along the inverse mean curvature flow and it plays a key role in the proof of the Riemannian Penrose inequality \cite{HI01}.

Consider the ingoing $(-)$ and outgoing $(+)$ null geodesic congruences from $\Sigma_t$. Let $\theta_t^{\pm}$ be the null expansions. Then the Hawking energy \cite{Ha68} of $\Sigma_t$ is defined as
\be\label{Hawkingenergy}
\mh(\Sigma_t)=\frac{1}{8\pi}\sqrt{\frac{\Area}{16\pi}}\int_{\Sigma_t} (\mbox{Scal}_{\sigma_t} +\theta_t^{+}\theta_t^{-})\d\sigma_t.\ee

Both $\mh(\Sigma)$ and $\eg(\Sigma)$ can be nonzero even in the flat Minkowski spacetime. This drawback is then corrected by Hayward \cite{Ha94}. By adding additional terms based on the double null foliation, his energy becomes zero for any generic two-surface in Minkowski spacetime. This motivates us to consider the following energy expression. Let $\theta_t^{\pm}$ and $(\sigma_t)_{ij}^{\pm}$ be the expansions and shear tensors of the congruences respectively.
\be\label{Haywardenergy}
\mhy(\Sigma)=\frac{1}{8\pi}\sqrt{\frac{|\Sigma_t|}{16\pi}}\int_\Sigma \Big(\mbox{Scal}_{\sigma_t} +\theta_t^{+}\theta_t^{-} -\frac{1}{2}(\sigma_t)_{AB}^{+}(\sigma_t)_{-}^{AB}\Big)\d\sigma_t.\ee

Note that the energy expression \eqref{Haywardenergy} is different from the original one Hayward suggested in \cite{Ha94} which contains an additional anoholonomicity term. In fact, the anoholonomicity is a boost-gauge-dependent quantity \cite[Page 61]{Sz09}. In this paper, we use the energy expression as Eqn (6.5) in \cite[Page 61]{Sz09} and still call it the Hayward energy.

Now we state our main theorem.
\begin{thm}
  Assume that the star has spatially constant matter density $\mu_0$. Let $\Sigma_t$ be a closed two-surface lying in a time slice $M_t=\{\mbox{t=const.}\}$ within the star. The surface is defined by an equation of the matter comoving coordinates $F(r,\Theta,\varphi)=0$.
  During the collapse, i.e. for $0\leq t < t^\ast=\sqrt{\frac{2}{3\mu_0}}$, we have the following conclusions:\\
  \noindent (i) The area of the surface $|\Sigma_t|$ is monotonically decreasing as $t$ increases. Indeed,
  \bee
  |\Sigma_t|=\Big(1-\sqrt{\frac{3\mu_0}{2}}\ t\Big)^{\frac{4}{3}}|\Sigma_0|.\eee
  \noindent (ii) The Geroch energy is nonpositive and it is monotonically increasing as $t$ increases. Indeed,
  \bee\begin{split} \eg(\Sigma_t)& = \Big(1-\sqrt{\frac{3\mu_0}{2}}\ t\Big)^{\frac{2}{3}}\sqrt{\frac{|\Sigma_0|}{16\pi}}\Big(\frac{\chi(\Sigma_0)}{2}-\frac{1}{16\pi} \int_{\Sigma_0}  H_0^2 \d \sigma_0 \Big)\\
  &=\Big(-\frac{1}{8\pi}\Big)\Big(1-\sqrt{\frac{3\mu_0}{2}}\ t\Big)^{\frac{2}{3}}\sqrt{\frac{|\Sigma_0|}{16\pi}}\int_{\Sigma_0}|\ringA_0|_{\sigma_0}^2\d \sigma_0.\end{split}\eee
   \noindent (iii) The Hawking energy is monotonically increasing as $t$ increases. Indeed,
  \bee
  \begin{split}
  \mh(\Sigma_t) &= \Big(1-\sqrt{\frac{3}{2}\mu_0}\ t\Big)^{\frac{2}{3}}\sqrt{\frac{|\Sigma_0|}{16\pi}}\Big(\frac{\chi(\Sigma_0)}{2}-\frac{1}{16\pi} \int_{\Sigma_0}  H_0^2 \d \sigma_0 \Big) + \frac{\mu_0}{24\pi^\frac{3}{2}}|\Sigma_0|^{\frac{3}{2}}\\
  &=\Big(-\frac{1}{8\pi}\Big)\Big(1-\sqrt{\frac{3\mu_0}{2}}\ t\Big)^{\frac{2}{3}}\sqrt{\frac{|\Sigma_0|}{16\pi}}\int_{\Sigma_0}|\ringA_0|_{\sigma_0}^2\d \sigma_0+ \frac{\mu_0}{24\pi^\frac{3}{2}}|\Sigma_0|^{\frac{3}{2}}.
  \end{split}\eee
\noindent (iv) The Hayward energy is conserved and remains positive. Indeed,
  \bee
  \mhy(\Sigma_t)=\frac{\mu_0}{24\pi^\frac{3}{2}}|\Sigma_0|^{\frac{3}{2}}.\eee
  Here $\chi(\Sigma_0)$ is the Euler-Poincar\'{e} characteristic number of the initial surface, $\sigma_0$, $|\Sigma_0|$, $H_0$, and $\ringA_0=A_0-\frac{H_0}{2}\sigma_0$ are the induced two-metric, the area, the mean curvature, and the traceless second fundamental form of the initial surface $\Sigma_0$ in the time slice $M_0=\{t=0\}$ respectively.
  \end{thm}
By taking the limit as $t\rightarrow t^{\ast-}$, we have the following corollary.
\begin{coro}
Let the notations be the same as in the main theorem above. Then we have
\bee
\lim_{t\rightarrow t^{\ast-}}|\Sigma_t|=0,\eee
\bee
  \lim_{t\rightarrow t^{\ast-}}\eg(\Sigma_t)=0,\eee
  and
  \bee
  \lim_{t\rightarrow t^{\ast-}}\mh(\Sigma_t)=\frac{\mu_0}{24\pi^\frac{3}{2}}|\Sigma_0|^{\frac{3}{2}}.\eee\end{coro}
We are ready to prove our main theorem.
\begin{proof}
 \noindent (i) $|\Sigma_t|=\Big(1-\sqrt{\frac{3\mu_0}{2}}\ t\Big)^{\frac{4}{3}}|\Sigma_0|$ immediately follows from
 \bee
 g_t=\Big(1-\sqrt{\frac{3\mu_0}{2}}\ t\Big)^{\frac{4}{3}}g_0\ \ \mbox{and} \ \ \d\sigma_t=\Big(1-\sqrt{\frac{3\mu_0}{2}}\ t\Big)^{\frac{4}{3}}\d \sigma_0.\eee
 \noindent (ii)
 \bee
\begin{split}\eg(\Sigma_t) &= \sqrt{\frac{\Area}{16\pi}}\Big(\frac{1}{8\pi}\int_{\Sigma_t}\mbox{Scal}_{\sigma_t}\d \sigma_t  - \frac{1}{16\pi} \int_{\Sigma_t} H_t^2 \d \sigma_t\Big)  \\
&=\Big(1-\sqrt{\frac{3\mu_0}{2}}\ t\Big)^{\frac{2}{3}}\sqrt{\frac{|\Sigma_0|}{16\pi}}\Big(\frac{\chi(\Sigma_t)}{2}-\frac{1}{16\pi} \int_{\Sigma_0} \big(e^{-\omega} H_0\big)^2e^{2\omega} \d \sigma_0 \Big)\\
&=\Big(1-\sqrt{\frac{3\mu_0}{2}}\ t\Big)^{\frac{2}{3}}\sqrt{\frac{|\Sigma_0|}{16\pi}}\Big(\frac{\chi(\Sigma_0)}{2}-\frac{1}{16\pi} \int_{\Sigma_0}  H_0^2 \d \sigma_0 \Big).
\end{split}\eee
where we have used the Gauss-Bonnet theorem and the fact that the Euler-Poincar\'{e} characteristic number remains unchanged.
According to \eqref{1+3}, the induced three-metric $g_t$ on the time slice $M_t$ is also flat. By the Gauss equation of $\Sigma_t$ in $M_t$, one has
\bee
\begin{split}
&\ \ \ R^{\sigma_t}(e_A,e_B,e_A,e_B)-A(e_A,e_A)A(e_B,e_B)+A(e_A,e_B)A(e_A,e_B)\\
&=R^{g_t}(e_A,e_B,e_A,e_B)\\
&=0.\end{split}\eee
Summing over $A,B=2,3$, it yields
\bee
\mbox{Scal}_{\sigma_t}-H_t^2+|A_t|_{\sigma_t}^2=\mbox{Scal}_{\sigma_t}-\frac{1}{2}H_t^2+|\ringA_t|_{\sigma_t}^2=0.\eee
Then one arrives at
\bee
\eg(\Sigma_t)=\Big(-\frac{1}{8\pi}\Big)\Big(1-\sqrt{\frac{3\mu_0}{2}}\ t\Big)^{\frac{2}{3}}\sqrt{\frac{|\Sigma_0|}{16\pi}}\int_{\Sigma_0}|\ringA_0|_{\sigma_0}^2\d \sigma_0.\eee
From this equation, it is obvious to see that the Geroch energy is nonpositive and is monotonically increasing.

\noindent (iii) By the Gauss equation of $\Sigma_t$ in spacetime, one can rewrite the Hawking energy \eqref{Hawkingenergy} as (cf. \cite[Page 59, Eqn (6.4)]{Sz09})
\bee
\begin{split}
\mh(\Sigma_t)
&=\frac{1}{8\pi}\sqrt{\frac{\Area}{16\pi}}\int_{\Sigma_t} \big(\mbox{Scal}_{\sigma_t} -\frac{1}{2}(H_t^2-p_t^2)\big)\d\sigma_t\\
&=\eg(\Sigma_t)+\frac{1}{16\pi}\sqrt{\frac{\Area}{16\pi}}\int_{\Sigma_t} p_t^2 \d \sigma_t.\end{split}\eee
Here $p_t=\mbox{tr}K_t|_{\Sigma_t}=K_t(e_2,e_2)+K_t(e_3,e_3)$.
Now
\bee\label{pterm}
\begin{split}
\frac{1}{16\pi}\sqrt{\frac{\Area}{16\pi}}\int_{\Sigma_t} p_t^2 \d \sigma_t &= \frac{e^{\omega}|\Sigma_0|^{\frac{1}{2}}}{16\pi\sqrt{16\pi}} \int_{\Sigma_0}\big(4\dot{\omega}^2  e^{2\omega}\big)\d \sigma_0\\
&=\frac{|\Sigma_0|^{\frac{1}{2}}}{16\pi^{\frac{3}{2}}}\big(\dot{\omega}^2 e^{3\omega}\big) \int_{\Sigma_0} \d \sigma_0\\
&=\frac{|\Sigma_0|^{\frac{3}{2}}}{16\pi^{\frac{3}{2}}} \frac{2\mu_0}{3}\\
&=\frac{\mu_0}{24\pi^\frac{3}{2}}|\Sigma_0|^{\frac{3}{2}}. \end{split}\eee
Here we have used \eqref{Kt} and \eqref{magic} and the monotonicity of the Hawking energy follows from that of the Geroch energy.

\noindent (iv) The contracted Gauss equation of $\Sigma_t$ in spacetime \cite[Page 833]{Ha94} reads
\bee
\mbox{Scal}_{\sigma_t} +\theta_t^{+}\theta_t^{-} -\frac{1}{2}(\sigma_t)_{AB}^{+}(\sigma_t)_{-}^{AB}=\sigma^{\alpha\mu}_t\sigma^{\beta\nu}_t R^{\tilde{g}}_{\alpha\beta\mu\nu}.\eee
In terms of the orthonormal frame $\{e_{\mu}\}$, the right hand side of the above equation turns out to be equal to $2R^{\tilde{g}}(e_2,e_3,e_2,e_3)$ - twice of the sectional curvature of the spacetime metric $\tilde{g}$ with respect to the plane spanned by $\{e_2,e_3\}$. Again, according to \eqref{Kt}, one has
\bee
\begin{split}
R^{\tilde{g}}(e_2,e_3,e_2,e_3)&=R^{g_t}(e_2,e_3,e_2,e_3)+K_t(e_2,e_2)K_t(e_3,e_3)-K_t(e_2,e_3)K_t(e_2,e_3)\\
&=0+\dot{\omega}^2-0\\
&=\dot{\omega}^2.
\end{split}
\eee
Then
\bee
\begin{split}
\mhy(\Sigma_t)&=\frac{1}{8\pi}\sqrt{\frac{|\Sigma_t|}{16\pi}}\int_{\Sigma_t} 2 R^{\tilde{g}}(e_2,e_3,e_2,e_3) \d \sigma_t\\
&=\frac{e^{\omega}|\Sigma_0|^{\frac{1}{2}}}{16\pi^{\frac{3}{2}}}\int_{\Sigma_0} \dot{\omega}^2 e^{2\omega}\d \sigma_0\\ &=\frac{|\Sigma_0|^{\frac{1}{2}}}{16\pi^{\frac{3}{2}}}\big(e^{\omega}\dot{\omega}^2 e^{2\omega}\big) \int_{\Sigma_0} \d \sigma_0\\ &=\frac{\mu_0}{24\pi^\frac{3}{2}}|\Sigma_0|^{\frac{3}{2}}.
\end{split}\eee
We have used \eqref{magic} in the last step. This completes the proof of the main theorem.
\end{proof}
We end this section by two concrete examples.
\begin{case} Let $\Sigma_t$ be a round sphere $\{r=a=\mbox{const.}\}$. Then \bee
|\Sigma_0|=4\pi a^2, \ H_0=\frac{2}{a}, \ |\ringA_0|_{\sigma_0}^2=0
\eee
and
\bee
\eg(\Sigma_t)=0, \
\mh(\Sigma_t)=\frac{\mu_0a^3}{3}, \
\mhy(\Sigma_t)=\frac{\mu_0a^3}{3}.
\eee
\end{case}
\begin{case}
  Let $\Sigma_t$ be a torus with major radius $a$ and minor radius $b$. It can parametrized as
  \bee\label{standardtoriiso}
\begin{split}
x^1&=(a+b\cos\theta)\cos\varphi,\\
x^2&=(a+b\cos\theta)\sin\varphi, \ \theta \in [0,2\pi), \ \varphi\in [0,2\pi)\\
x^3&=b\sin\theta.
\end{split}\eee
Here $\{x^1,x^2,x^3\}$ are the matter comoving Cartesian coordinates and $r=\sqrt{(x^1)^2+(x^2)^2+(x^3)^2}$. The induced two-metric becomes \bee
\sigma_0=b^2\d \theta^2+(a+b\cos\theta)^2 \d \varphi^2.
\eee
Further calculation shows
\bee
|\Sigma_0|=4\pi^2 a b, \ H_0=\frac{a+2b\cos\theta}{b(a+b\cos\theta)}, \ |\ringA_0|_{\sigma_0}^2=\frac{a^2}{2b^2(a+b\cos\theta)^2}.
\eee
Then
\bee
\begin{split}
\eg(\Sigma_t)&=-\frac{1}{8\pi}\Big(1-\sqrt{\frac{3}{2}\mu_0}\ t\Big)^{\frac{2}{3}}\sqrt{\frac{4\pi^2ab}{16\pi}}\Big(\int_{0}^{2\pi}\d \varphi \Big)\Big(\int_0^{2\pi}
\frac{a^2}{2b(a+b\cos\theta)}\d \theta\Big)\\
&=-\frac{a^{\frac{5}{2}}\pi^{\frac{3}{2}}}{8\sqrt{b(a^2-b^2)}}
\Big(1-\sqrt{\frac{3}{2}\mu_0}\ t\Big)^{\frac{2}{3}},
\end{split}
\eee
\bee
\begin{split}
\mh(\Sigma_t)&=-\frac{\sqrt{ab\pi}}{16}\Big(1-\sqrt{\frac{3}{2}\mu_0}\ t\Big)^{\frac{2}{3}}\int_0^{2\pi}\frac{(a+2b\cos\theta)^2}{b(a+b\cos\theta)}\d \theta+\frac{\mu_0(ab\pi)^{\frac{3}{2}}}{3}\\
&=-\frac{a^{\frac{5}{2}}\pi^{\frac{3}{2}}}{8\sqrt{b(a^2-b^2)}}
\Big(1-\sqrt{\frac{3}{2}\mu_0}\ t\Big)^{\frac{2}{3}}+
\frac{\mu_0(ab\pi)^{\frac{3}{2}}}{3},\\
\mhy(\Sigma_t)&=\frac{\mu_0(ab\pi)^{\frac{3}{2}}}{3}.
\end{split}
\eee
For $a=0.5$, $b=0.4$ and $\mu_0=1$, the graphs of $\eg(\Sigma_t)$, $\mh(\Sigma_t)$ and
$\mhy(\Sigma_t)$ of $\Sigma_t$ with respect to $t$ are plotted in Fig.\ref{3typeenergy}.

\begin{figure}[htp]
\begin{center}
\includegraphics[width=0.50\textwidth]{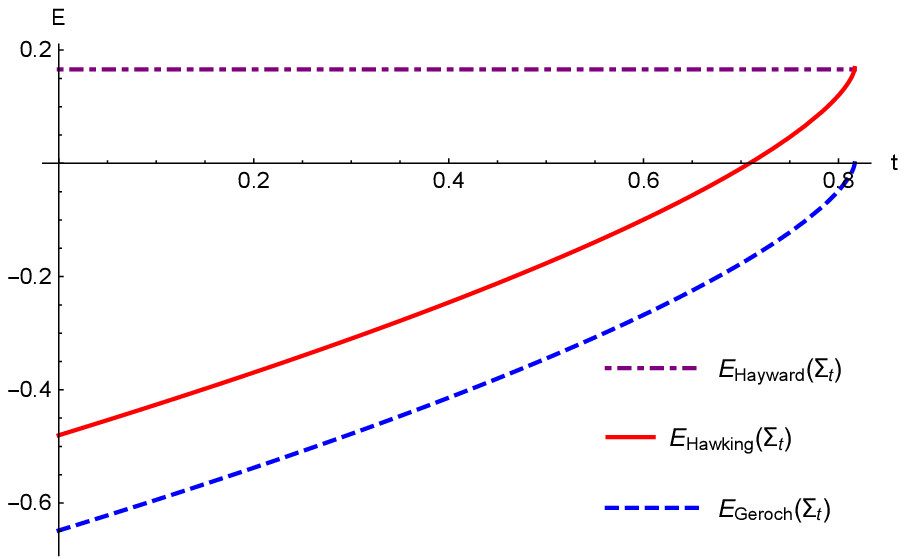}
\caption{\label{3typeenergy} Graphs of $\eg(\Sigma_t)$, $\mh(\Sigma_t)$ and
$\mhy(\Sigma_t)$ of $\Sigma_t$ with respect to $t$ for $a=0.5$, $b=0.4$ and $\mu_0=1$. }
\end{center}
\end{figure}

\end{case}

\section{Conclusions and Discussions}\label{S4}
We consider the gravitational collapse of the Oppenheimer-Snyder dust cloud. The initial matter density $\mu_0$ is assumed to be spatially constant. For a generic closed two-surface within the star lying in a time slice with an defining equation of the matter comoving coordinates, three quasi-local energies are investigated. The Geroch energy of the surface is nonpositive and increases to zero. The Hawking energy of the surface can start from a negative value and increases to the Hayward energy of the initial surface. The Hayward energy behaves quite well. The preference appears to be supported by the fact that it is conserved and remains positive during the collapse. It should be emphasized that even though the spacetime is spherically symmetric, there are no restrictions on the topology and the symmetry of the surface in our results.

There are also works for non-homogenous dust clouds in the literature \cite{Gu73,Hu74,MSY73,MSY74,Chr84,N86}. If one relaxes the condition of the matter density from constant to being monotonically decreasing as $r$ increases, it has been shown that the shell with increasing $r$ arrives successively at the centre and there is no shell-crossing \cite{Gu73}. In this case,
\bee
e^{\omega(t,r)}=R^\prime(t,r)=\Big(r^{\frac{3}{2}}-h^{\frac{1}{2}}(r)t\Big)^{-\frac{1}{3}}\Big(r^{\frac{1}{2}}-\frac{1}{3}h^{-\frac{1}{2}}(r)h^{\prime}(r)t\Big)\eee
and
\bee
R(t,r)=\Big(r^{\frac{3}{2}}-h^{\frac{1}{2}}(r)t\Big)^{\frac{2}{3}}
\eee
where $h^{\frac{1}{2}}(r)=\frac{3}{2}\sqrt{2M(r)}$ and $M(r)$ is defined in \eqref{mass} (cf. \cite[Page 100, Eqn (12.16)]{Ch09}).
The spacetime metric \eqref{spme} is not sliced with flat three-spaces as in \eqref{1+3} and further \eqref{magic} and \eqref{Kt} are no longer valid which play key roles in the proof of our main theorem - the blow-up of the curvature being neutralized by the shrinking of the area form appropriately. Moreover, although in the new coordinates $(t,R,\Theta,\varphi)$, the induced three-metric on the time slice still reduces to the Euclidean metric, nonzero component in front of the cross-term $\d t \d R$ appears \cite[Page 101, Eqn (12.19)]{Ch09},
\bee
\tilde{g}=-\big(1-\frac{\sqrt{2M(r)}}{R}\big)\d t^2+2R^{-\frac{1}{2}}\sqrt{2M(r)}\d R \d t  +\d R^2 +R^2\d \Theta^2 +R^2\sin^2\Theta \d \varphi^2. \eee
Then the time slice is no longer totally umbilical. This also brings a peck of difficulties and leads to the consequence that the quasi-local energy behaves badly.

When the initial matter density is not monotonically decreasing, the scenario is much more complicated. Shell-crossing happens and it inevitably leads to the non-central singularities \cite{Gu73}. The study of the behaviour of the quasi-local energy becomes significantly more formidable. Clearly it is beyond the scope of the current paper.

\section*{Acknowledgments}
 The authors would like to thank the referees for very useful comments and suggestions. X. He is partially supported by the Natural Science Foundation of Hunan Province (Grant 2018JJ2073). N. Xie is partially supported by the National Natural Science Foundation of China (Grant 11671089).


\end{document}